# Power Allocation Strategies for Fixed-Gain Half-Duplex Amplify-and-Forward Relaying in Nakagami-m Fading

Ammar Zafar, *Student Member, IEEE*, Redha M. Radaydeh, *Member IEEE*, Yunfei Chen, *Senior Member, IEEE*, and Mohamed-Slim Alouini, *Fellow, IEEE*

**Abstract**

In this paper, we study power allocation strategies for a fixed-gain amplify-and-forward relay network employing multiple relays. We consider two optimization problems for the relay network: 1) optimal power allocation to maximize the end-to-end signal-to-noise ratio (SNR) and 2) minimizing the total consumed power while maintaining the end-to-end SNR over a threshold value. We investigate these two problems for two relaying protocols of all-participate relaying and selective relaying and multiple cases of available channel state information (CSI) at the relays. We show that the SNR maximization problem is concave and the power minimization problem is convex for all protocols and CSI cases considered. We obtain closed-form expressions for the two problems in the case for full CSI and CSI of all the relay-destination links at the relays and solve the problems through convex programming when full CSI or CSI of the relay-destination links are not available at the relays. Numerical results show the benefit of having full CSI at the relays for both optimization problems. However, they also show that CSI overhead can be reduced by having only partial CSI at the relays with only a small degradation in performance.

This work was funded by King Abdullah University of Science and Technology (KAUST).

A. Zafar and M. -S. Alouini are with the Computer, Electrical and Mathematical Sciences & Engineering (CEMSE) Division, King Abdullah University of Science and Technology (KAUST), Thuwal, Makkah Province, Saudi Arabia. 23955-6900 (E-mail: {ammar.zafar, slim.alouini}@kaust.edu.sa).

Y. Chen is with the School of Engineering, University of Warwick, Coventry, UK. CV4 7AL (E-mail: yunfei.chen@warwick.ac.uk).

R. M. Radaydeh was with KAUST. He is now with the Electrical Engineering Department, Alfaisal University, Riyadh, Saudi Arabia. 11533 (E-mail: redha.radaydeh@gmail.com).

# I. INTRODUCTION

The classical three-terminal relay channel has been around since the 1960s when it was first introduced by Van Der Meulen [1], [2]. Cover and Gamal then characterized the capacity of a three-terminal relay channel where a relay assists a source node in communicating with a destination node [3]. Since then precious little work was carried out on relays until a decade ago. The rapid progress in wireless communications technology, the increasing popularity of tetherless connectivity and satisfying the high quality-of-service (QoS) requirements have rekindled interest in relays [4], [5]. Recent years have seen a dearth of work being carried out to study the performance of relay-assisted systems [6]–[13]. It has been shown that relays can provide spatial diversity [14]–[16], increase capacity [17]–[19], conserve power [20], [21] and enhance coverage [22]–[24].

There are two main protocols for relays: 1) amplify-and-forward (AF) (non-regenerative) relays, 2) decode-and-forward (DF) (regenerative relays) [25]. AF relays amplify the received signal and forward it to the intended destination. AF relays are unable to detect any errors in the received signal and hence, are termed as non-regenerative. Unlike AF relays, DF relays can detect the presence of an error and can correct it before re-transmission and hence, are termed as regenerative. However, DF relays require extra processing that increases complexity at the relays which often needs to be simple. In the following, we consider AF relays which can be further divided into two groups; namely, fixed-gain AF relays and variable-gain AF relays [26]. In fixed-gain AF relays, the relay gain does not depend on the fading gain of the source-relay link and remains constant throughout the duration of system operation, whereas in variable-gain AF relays, the relay gain changes with the fading gain of the source-relay link. Thus, in variable-gain relays, the relay always requires CSI of the source-relay link. In this work, we concentrate on fixed-gain AF relays.

Power is a precious resource. It was reported in [27] that information and communication technology (ICT) utilizes more than $3\%$ of the total electrical energy consumed worldwide and this percentage is expected to increase with time. Furthermore, in wireless communications, vendors and operators are searching for energy-efficient algorithms and devices to cut down on energy and operating costs [28]. In addition, mobile devices need to conserve energy as they have limited battery. Therefore, it is of paramount importance to use the available power as efficiently as possible. In a cooperative relay network, this corresponds to allocating power efficiently among the source and relay nodes to improve performance. Moreover, it is also essential to reduce overhead in the system. Most of the works available on power allocation for relay networks focus on the case of full CSI available at the relays. This usually requires the controller [1] to feedback CSI of all the links to each relay. This causes great overhead and consumes precious system resources such as bandwidth and time. Thus, it is crucial to come up with power allocation strategies which can work with partial CSI or channel statistics at the relay to minimize the feedback overhead [29]. However, this reduced overhead comes at the cost of decrease in performance as there is less knowledge to work with and exploit. Therefore, there is a performance-overhead trade-off. It is good to have insight into this trade-off, so that optimal decisions can be taken to improve system performance under different scenarios.

Power allocation for cooperative network with fixed-gain AF relays was studied in [30]–[34][2]. In [30], the authors proposed power allocation schemes to maximize the sum and product of the average SNR of the source-destination and average SNR of the relay-destination link for a fixed gain relay-assisted source-destination pair. Moreover, the schemes required only the knowledge of channel statistics. For the same model as in [30], [31] proposed optimal and near-optimal power allocation algorithms to maximize the end-to-end SNR under slowly varying channel conditions. In [32], power allocation to minimize the symbol error rate (SER) for $M$-ary phase shift keying (MPSK) was derived for a source communicating with destination through a single fixed-gain AF relay. It was shown in [32] that the power allocation method provided better performance when the relay was near the destination. References [33] and [34], both proposed power allocation strategies to minimize the bit error rate (BER) of a communication system employing a single fixed-gain AF relay. In [35], we studied power allocations for an AP fixed gain AF relay network with complete CSI of the relay-destination link and knowledge of only channel statistics of the source-destination and source-relay links. We provided closed-form solutions for the optimization problems considered[3].

In this work, we consider a source-destination pair which communicates with the help of $m$ fixed-gain AF relays. In addition to the $m$ dual-hop links, the source and destination are also connected through a direct link[4]. The relays are assumed to have only one antenna and work in half-duplex mode. Thus, they cannot transmit and receive at the same time and at the same frequency. To avoid interference, all the relays are assumed to operate on orthogonal channels[5]. For this system, we consider two relay participation schemes of all-participate (AP) relaying in which all the $m$ relays forward the signal to the destination and selective relaying in which only the selected[6] relay forwards the signal received from the source to the destination. For both schemes, we consider two optimization problems of optimal power allocation (OPA) and energy-efficiency. We refer to these as dual problems. OPA refers to the problem of allocating power to the source and the relays to maximize the end-to-end SNR under a total power constraint on the system. In the dual problem of energy-efficiency, the total power consumed is minimized while keeping the end-to-end SNR above a certain threshold. Furthermore, we consider three cases of available CSI at the relays. In all the three cases, it is assumed that the destination has full CSI. The three cases of CSI considered are:

1) Full CSI of all the links at the relays.
2) Knowledge of the channel statistics of all the links at the relays.
3) Two cases of partial CSI at the relays
   a) Full CSI of all the relay-destination links and knowledge of only channel statistics of all the source-relay and source-destination links (Partial CSI-$\beta$).
   b) Full CSI of all the source-relay and source-destination links and knowledge of only channel statistics of all the relay-destination links (Partial CSI-$\alpha$).

---

[1] The place where system decisions are taken, for instance it can be the destination.

[2] There are many other works on power allocation for relays, however most of them focus on variable-gain relays.

[3] We made a slight mistake for the implementation of the energy-efficiency problem solution in [35] which makes the results presented in [35] a little worse than they actually should be. This mistake and correction will be discussed in Section II-C1

We study the two dual optimization problems for the two relay participation schemes under all three cases of CSI given above.

Our main contribution in this paper is to give a detailed analysis of the system under AP relaying, selective relaying and the three CSI cases. We give insight on the performance-overhead trade-off. We show which system configurations perform better for which objectives under which conditions. For instance, numerical results show that when the fading gains of the source-relay links and the source-destination links are the same on average, then having knowledge of the instantaneous CSI of the source-relay links

---

[4]The results in this paper include the case of no direct link as a special scenario by replacing the fading gain of the source-destination link by 0.

[5]They can be orthogonal in time, frequency, space, or code.

[6]How selection takes place is discussed in Section III when we consider selective relaying in detail.

benefits more than having knowledge of the source-destination links for OPA.

The remainder of the paper is organized as follows. In Section II, we consider AP relaying. We discuss both optimization problems of OPA and energy-efficiency for all the CSI cases listed above. Section III focuses on selective relaying. We again discuss OPA and energy-efficiency under all three assumptions on the CSI. Numerical results are presented in Section IV. Finally, Section V summarizes the main results of the paper.

## II. ALL-PARTICIPATE SCHEME

### A. System Model

Consider a cooperative system with a source node, a destination node, and $m$ relays. Each relay is assumed to be equipped with only one antenna and operates in half-duplex mode. The source and the relays transmit on orthogonal channels. Without loss of generality, we assume time division multiple access (TDMA). The transmission takes place in two phases. In the first phase, the source transmits information to the relays and the destination. The signals at the $i$th relay and the destination are given by

$$y_{si} = \sqrt{E_s}h_{si}s + n_{si} \tag{1}$$

$$y_{sd} = \sqrt{E_s}h_{sd}s + n_{sd}, \tag{2}$$

where $E_s$ is the source energy, $h_{si}$ and $h_{sd}$ are the instantaneous channel gains from the source to the $i$th relay and destination respectively. The channels are modeled as independent Nakagami-m random variables. $n_{si}$ and $n_{id}$ are the complex Gaussian noise with zero mean and variances $\sigma_{si}^2$ and $\sigma_{sd}^2$, respectively. In the second phase, the relays, after amplification, forward the signal to the destination. The received signal at the destination from the $i$th relay is

$$y_{id} = \sqrt{a_i E_s E_i}h_{si}h_{id}s + n_i, \tag{3}$$

where $a_i$ is the $i$th relay gain, $E_i$ is the $i$th relay energy, $h_{id}$ is the instantaneous channel gain from the $i$th relay to the destination which is again modeled as Nakagami-m fading, $n_i$ is $n_i \sim CN(0, \sigma_i^2)$ and

$$\sigma_i^2 = a_i E_i |h_{id}|^2 \sigma_{si}^2 + \sigma_{id}^2.$$

One can write the $m+1$ received signals at the destination in matrix form

$$\mathbf{y} = \mathbf{h}s + \mathbf{n}, \tag{4}$$

where

$$\mathbf{y} = \left[\frac{1}{\sigma_{sd}}y_{sd} \ \ \frac{1}{\sigma_1}y_{1d} \ldots \frac{1}{\sigma_m}y_{md}\right]^T,$$

$$\mathbf{h} = \left[\sqrt{\frac{E_s}{\sigma_{sd}^2}}h_{sd} \ \sqrt{\frac{a_i E_s E_1}{a_i E_1 |h_{1d}|^2 \sigma_{s1}^2 + \sigma_{1d}^2}}h_{s1}h_{1d} \ldots \sqrt{\frac{a_i E_t E_m}{a_i E_m |h_{md}|^2 \sigma_{sm}^2 + \sigma_{md}^2}}h_{sm}h_{md}\right]^T,$$

and $\mathbf{n} \sim CN(\mathbf{0}, \mathbf{I})$. Throughout this paper, it is assumed that the destination has complete CSI of all the links. It is also assumed that all the links experience independent fading. Furthermore, the fading gain of each link changes independently from one time slot to another.

## B. Optimal Power Allocation

First, we consider the problem of OPA. In OPA, the end-to-end SNR is maximized under power constraints on the system. For OPA, we consider the three cases of full CSI at the relays, knowledge of only channel statistics at the relays and partial CSI at the relays.

*1) Full CSI:* In this section, we assume that the relays also have complete CSI of all the links. Using (4), and assuming maximal-ratio-combining (MRC) at the destination, the end-to-end SNR of the system is given by

$$\gamma = E_s \left( \sum_{i=0}^{m} \alpha_i - \sum_{i=1}^{m} \frac{\alpha_i \zeta_i}{a_i E_i \beta_i + \zeta_i} \right), \tag{5}$$

where $\alpha_0 = \frac{|h_{sd}|^2}{\sigma_{sd}^2}$, $\alpha_i = \frac{|h_{si}|^2}{\sigma_{si}^2}$, $\beta_i = \frac{|h_{id}|^2}{\sigma_{id}^2}$ and $\zeta_i = \frac{1}{\sigma_{si}^2}$. In this work, we consider a total power constraint on the whole system and individual power constraints on all the nodes. Hence, the power allocation problem can be written as

$$\max_{E_s, E_i} \gamma, \text{ subject to}$$
$$E_s + \sum_{i=1}^{m} E_i \leq E_{tot}, \ 0 \leq E_s \leq E_s^{max}, \ 0 \leq E_i \leq E_i^{max}. \tag{6}$$

In general, $\gamma$ is not a concave function of the source and relay powers as its Hessian is not negative semi-definite. However, as we show in Appendix A, the objective function in (6) is concave given the constraints on the system. Hence, the optimization problem, (6), can be solved using the Lagrange dual method [36]. Moreover, as the constraints are affine, Slater's condition [36] is satisfied, i.e. the duality gap between the primal and dual solutions is zero. Therefore, the solution obtained for the Lagrange dual problem is also the optimal solution of the primal problem in (6). Hence, solving the problem in (6) using the Lagrange dual method, with the help of [35], yields the solution

$$E_s = \left( \frac{\delta \left( \sum_{i=1}^{m} \sqrt{\frac{\alpha_i \zeta_i}{a_i \beta_i}} \right)^2}{\left( \sum_{i=0}^{m} \alpha_i - \delta \right)^2} \right)_0^{E_s^{max}} \tag{7}$$

$$E_j = \left( \frac{\left( \sum_{i=1}^{m} \sqrt{\frac{\alpha_i \zeta_i}{a_i \beta_i}} \right)}{\left( \sum_{i=0}^{m} \alpha_i - \delta \right)} \sqrt{\frac{\alpha_j \zeta_j}{a_j \beta_j}} - \frac{\zeta_j}{a_j \beta_j} \right)_0^{E_j^{max}}, \tag{8}$$

where

$$\delta = \sum_{i=0}^{m} \alpha_i - \left( \sum_{i=1}^{m} \sqrt{\frac{\alpha_i \zeta_i}{a_i \beta_i}} \right) \sqrt{\frac{\left( \sum_{i=0}^{m} \alpha_i \right)}{E_{tot} + \sum_{j=1}^{m} \frac{\zeta_j}{a_j \beta_j}}}. \tag{9}$$

From equations (7) and (8), one can conclude that the OPA follows a water-filling solution. Hence, power is allocated in an iterative manner. However, unlike traditional water-filling algorithm, here the closed-form solution may change according to the initial results. If the source power, $E_s$, satisfies its constraints and any relay power does not satisfy its individual constraint, then the solution to the problem remains the same, however, the optimization variables and the constraint changes. Therefore, in this case, the optimal solution is given by

$$E_s = \left( \frac{\delta \left( \sum_{i \in \text{X}} \sqrt{\frac{\alpha_i \zeta_i}{a_i \beta_i}} \right)^2}{\left( \sum_{i \notin \text{Y}} \alpha_i - \delta - \sum_{i \in \text{Z}} \frac{\alpha_i \zeta_i}{a_i E_i^{max} \beta_i + \zeta_i} \right)^2} \right)_0^{E_s^{max}} \tag{10}$$

$$E_j = \left( \frac{\left( \sum_{i \in \text{X}} \sqrt{\frac{\alpha_i \zeta_i}{a_i \beta_i}} \right) \sqrt{\frac{\alpha_j \zeta_j}{a_j \beta_j}}}{\sum_{i \notin \text{Y}} \alpha_i - \delta - \sum_{i \in \text{Z}} \frac{\alpha_i \zeta_i}{a_i E_i^{max} \beta_i + \zeta_i}} - \frac{\zeta_j}{a_j \beta_j} \right)_0^{E_j^{max}} \tag{11}$$

and

$$\delta = \sum_{i \notin Y} \alpha_i - \sum_{i \in Z} \frac{\alpha_i \zeta_i}{a_i E_i^{max} \beta_i + \zeta_i} - \left( \sum_{i \in X} \sqrt{\frac{\alpha_i \zeta_i}{a_i \beta_i}} \right) \sqrt{\frac{\sum_{i \notin Y} \alpha_i - \sum_{i \in Z} \frac{\alpha_i \zeta_i}{a_i E_i^{max} \beta_i + \zeta_i}}{E_{tot} - \sum_{i \in Z} E_i^{max} + \sum_{j \in X} \frac{\zeta_j}{a_j \beta_j}}},$$

where X, Y and Z represent the sets of powers which satisfy the individual constraints, are less than zero and greater than the peak individual constraint, respectively. Now, if the source power is greater than $E_s^{max}$, then it is set at $E_s^{max}$. The updated optimal solution for the relay powers now becomes

$$E_j = \left( \sqrt{\frac{E_s^{max} \alpha_j \zeta_j}{\delta a_j \beta_j}} - \frac{\zeta_j}{a_j \beta_j} \right)_0^{E_j^{max}}, \quad (12)$$

where the Lagrangian multiplier can be obtained as

$$\delta = \frac{E_s^{max} \left( \sum_{j \in X} \sqrt{\frac{\alpha_j \zeta_j}{a_j \beta_j}} \right)^2}{\left( E_{tot} - E_s^{max} - \sum_{i \in Z} E_i^{max} + \sum_{j \in X} \frac{\zeta_j}{a_j \beta_j} \right)^2}.$$

*2) Knowledge of only Channel Statistics:* In Section II-B1, it was assumed that the each relay had full CSI of all the links. However, this greatly increases the complexity of the system. The destination has to inform all relays of the CSI through dedicated feedback channels. This consumes a considerable amount of resources. Moreover, the reverse link between the destination and the relays might be poor and communication might not be possible. Therefore, it is desirable to be able to work with less CSI. Hence, in this section, we assume that each relay only has knowledge of the channel statistics of all the links and the destination has full CSI. Thus, to perform power allocation, the end-to-end SNR needs to be averaged over all the links. As the channels are modeled as Nakagami-m random variable, $\alpha_i$ and $\beta_i$ are both Gamma random variables and their probability density functions and are given by

$$f_{\alpha_i}(x) = \frac{1}{\Gamma(k_{\alpha_i}) \bar{\gamma}_{\alpha_i}^{k_{\alpha_i}}} x^{k_{\alpha_i}-1} e^{-\frac{x}{\bar{\gamma}_{\alpha_i}}} \quad \text{and} \quad f_{\beta_i}(x) = \frac{1}{\Gamma(k_{\beta_i}) \bar{\gamma}_{\beta_i}^{k_{\beta_i}}} x^{k_{\beta_i}-1} e^{-\frac{x}{\bar{\gamma}_{\beta_i}}} \quad x \geq 0, \quad (13)$$

respectively, where $k_{\alpha_i}$ and $k_{\beta_i}$ are the shape parameters of the links, $\bar{\gamma}_{\alpha_i}$ and $\bar{\gamma}_{\beta_i}$ are the average SNRs of the links and $\Gamma(.)$ is the gamma function [37, Eq. (8.310.1)] As all the links are assumed to be independent, the average end-to-end SNR can be found by averaging (5) over the density functions in (13)

$$\bar{\gamma} = \int_0^\infty \cdots \int_0^\infty E_s \left( \sum_{i=0}^m x_i - \sum_{i=1}^m \frac{x_i \zeta_i}{a_i E_i y_i + \zeta_i} \right) \frac{1}{\Gamma(k_{\alpha_0})\bar{\gamma}_{\alpha_0}^{k_{\alpha_0}}} x^{k_{\alpha_0}-1} e^{-\frac{x}{\bar{\gamma}_{\alpha_0}}} \times \\ \prod_{i=1}^m \frac{1}{\Gamma(k_{\alpha_i})\bar{\gamma}_{\alpha_i}^{k_{\alpha_i}}} x_i^{k_{\alpha_i}-1} e^{-\frac{x_i}{\bar{\gamma}_{\alpha_i}}} \frac{1}{\Gamma(k_{\beta_i})\bar{\gamma}_{\beta_i}^{k_{\beta_i}}} y_i^{k_{\beta_i}-1} e^{-\frac{y_i}{\bar{\gamma}_{\beta_i}}} dx_i dy_i \quad (14)$$

Solving the above with the help of [37, Eq. (3.383.10)] gives the average end-to-end SNR as

$$\bar{\gamma} = E_s \sum_{i=0}^m k_{\alpha_i} \bar{\gamma}_{\alpha_i} - E_s \sum_{i=1}^m \frac{k_{\alpha_i} \bar{\gamma}_{\alpha_i} \zeta_i^{k_{\beta_i}}}{a_i^{k_{\beta_i}} \bar{\gamma}_{\beta_i}^{k_{\beta_i}}} \frac{1}{E_i^{k_{\beta_i}}} e^{\frac{\zeta_i}{a_i E_i \bar{\gamma}_{\beta_i}}} \Gamma\left(1 - k_{\beta_i}, \frac{\zeta_i}{a_i E_i \bar{\gamma}_{\beta_i}}\right) \quad (15)$$

where $\Gamma(.,.)$ is the upper incomplete gamma function given by $\Gamma(s,z) = \int_z^\infty e^{-t} t^{s-1} dt$ [37, Eq. (8.350.2)]. The optimization problem is the same as in II-B1, however the objective function is now changed to $\bar{\gamma}$ instead of $\gamma$. We show that $\bar{\gamma}$ is a concave function of the optimization parameters on the domain of interest in Appendix B. Thus, the optimization problem is concave. However, it is difficult to find closed-form expressions for the optimal solution due to the complexity of the objective function. Fortunately, as the problem is concave, we can utilize well-known algorithms for convex optimization. So, one such algorithm, the interior point algorithm can be used to find the optimal solution [36].

For the special case of Rayleigh fading, $\bar{\gamma}$ in (15) simplifies to

$$\bar{\gamma} = E_s \sum_{i=0}^m \bar{\gamma}_{\alpha_i} - E_s \sum_{i=1}^m \frac{\zeta_i \bar{\gamma}_{\alpha_i}}{a_i E_i \bar{\gamma}_{\beta_i}} e^{\frac{\zeta_i}{a_i E_i \bar{\gamma}_{\beta_i}}} E_1\left(\frac{\zeta_i}{a_i E_i \bar{\gamma}_{\beta_i}}\right), \quad (16)$$

where $E_1(.)$ is the exponential integral function of the first kind and is related to the exponential integral function as $E_1(x) = -E_i(-x)$ [38].

*3) Partial CSI:* Two important sub-cases of the full CSI and knowledge of channel statistics cases discussed in the previous two sections are the partial CSI instances: 1) partial CSI-$\beta$ 2) partial CSI-$\alpha$. These are important as they give us an idea regarding the trade-off between performance and complexity. Studying the performance of these two cases of partial CSI provides us insight as to knowledge of which link is more important. Thus, if an increase in performance is desired, then only the links which have greater affect on the performance of the system can be fed back to the relays.

For the first case, the objective function can be obtained from $\gamma$ by replacing $\alpha_i$ by $k_{\alpha_i} \bar{\gamma}_{\alpha_i}$. Then power allocation is done according to the algorithm described in Section II-B1. In the second case, the objective function can be obtained from $\bar{\gamma}$ by replacing $k_{\alpha_i}\bar{\gamma}_{\alpha_i}$ by $\alpha_i$. Then resource allocation can be performed using the interior-point algorithm as in Section II-B2.

## C. Energy-Efficiency

In Section II-B, we considered the problem of maximizing the end-to-end SNR under individual and total power constraints. In this section, we study the problem of energy-efficiency. The objective is to minimize the total power consumed, $E_{tot} = E_s + \sum_{i=1}^{m} E_i$, while keeping the instantaneous end-to-end SNR above a threshold, $\gamma^{th}$, and ensuring the source and relay powers do not exceed their respective individual constraints. Like Section II-B, we consider the three cases of full CSI, knowledge of only channel statistics and partial CSI.

*1) Full CSI:* With full CSI at the relay, the optimization problem is given by

$$\min_{E_s, E_i} E_{tot}, \text{ subject to}$$
$$\gamma \geq \gamma^{th}, \ 0 \leq E_s \leq E_s^{max}, \ 0 \leq E_i \leq E_i^{max}. \quad (17)$$

Now, we need to show that the optimization problem in (17) is convex and hence, the optimal solution can be found. The objective function and the individual constraints are convex function. So, it is only left to show that $\gamma$ is concave, meaning $\gamma^{th} - \gamma$ is convex, on the domain of the problem. Employing the same notation and procedure as in Appendix A, to prove concavity of $\gamma$, we need to show that $D(x, y) = (f(x) - f(y))(g(x) - g(y)) \leq 0$. If $f(x) > f(y)$, then $g(y) > g(x)$, to satisfy the constraint on $\gamma$ and vice versa. Thus, $D(x, y) < 0$ and $\gamma$ is concave on the domain. Moreover, $\gamma$ is a monotonically increasing function of $E_s$ and $E_i$s, hence the optimal solution to (17) is achieved when $\gamma = \gamma^{th}$. As the other two constraints are affine and the objective function is convex, Slater's condition is satisfied. Therefore, the solution obtained using the Lagrange dual method will be optimal. Solving the problem (17) using the Lagrange dual method, with the help of [35], gives the optimal solution as

$$E_s = \left( \frac{\rho \left( \sum_{j=1}^{m} \sqrt{\frac{\alpha_j \zeta_j}{a_j \beta_j}} \right)^2}{(\rho \sum_{i=0}^{m} \alpha_i - 1)^2} \right)_0^{E_s^{max}} \quad (18)$$

$$E_j = \left( \frac{\rho \left( \sum_{i=1}^{m} \sqrt{\frac{\alpha_i \zeta_i}{a_i \beta_i}} \right)}{(\rho \sum_{i=0}^{m} \alpha_i - 1)} \sqrt{\frac{\alpha_j \zeta_j}{a_j \beta_j}} - \frac{\zeta_j}{a_j \beta_j} \right)_0^{E_j^{max}}, \quad (19)$$

where

$$\rho = \frac{\left(\sum_{j=1}^{m} \sqrt{\frac{\alpha_j \zeta_j}{a_j \beta_j}}\right)}{\sum_{i=0}^{m} \alpha_i \sqrt{\gamma^{th}}} + \frac{1}{\sum_{i=0}^{m} \alpha_i}. \quad (20)$$

From (18) and (19), one can see that the solution again follows a water-filling algorithm described in Section II-B1. Hence, the power allocation process is repeated until all the powers satisfy the constraint. However, the optimal solution changes depending on the initial power allocation. There are two cases like for the problem of OPA: 1) $E_s$ lies between 0 ans $E_s^{max}$, 2) $E_s$ is greater than $E_s^{max}$. Considering case 1 first, the optimal power allocation is given by

$$E_s = \left(\frac{\rho \left(\sum_{j \in \mathrm{X}} \sqrt{\frac{\alpha_j \zeta_j}{a_j \beta_j}}\right)^2}{\left(\rho \sum_{i \notin \mathrm{Y}} \alpha_i - \rho \sum_{i \in \mathrm{Z}} \frac{\alpha_i \zeta_i}{a_i E_i^{max} \beta_i + \zeta_i} - 1\right)^2}\right)_0^{E_s^{max}}$$

$$E_j = \left(\frac{\rho \left(\sum_{i \in \mathrm{X}} \sqrt{\frac{\alpha_i \zeta_i}{a_i \beta_i}}\right) \sqrt{\frac{\alpha_j \zeta_j}{a_j \beta_j}}}{\rho \sum_{i \notin \mathrm{Y}} \alpha_i - \rho \sum_{i \in \mathrm{Z}} \frac{\alpha_i \zeta_i}{a_i E_i^{max} \beta_i + \zeta_i} - 1} - \frac{\zeta_j}{a_j \beta_j}\right)_0^{E_i^{max}}$$

and

$$\rho = \frac{\sqrt{\frac{\left(\sum_{j \in \mathrm{X}} \sqrt{\frac{\alpha_j \zeta_j}{a_j \beta_j}}\right)^2}{\gamma^{th}}} + 1}{\sum_{i \notin \mathrm{Y}} \alpha_i - \sum_{i \in \mathrm{Z}} \frac{\alpha_i \zeta_i}{a_i E_i^{max} \beta_i + \zeta_i}}.$$

In the second case, $E_s = E_s^{max}$ and

$$E_j = \left(\sqrt{\frac{\rho \alpha_j \zeta_j E_s^{max}}{a_j \beta_j}} - \frac{\zeta_j}{a_j \beta_j}\right)_0^{E_j^{max}}, \quad (21)$$

$$\rho = \frac{E_s^{max} \left(\sum_{i \in \mathrm{X}} \sqrt{\frac{\alpha_i \zeta_i}{a_i \beta_i}}\right)^2}{\left(E_s^{max} \left(\sum_{i \notin \mathrm{Y}} \alpha_i - \sum_{i \in \mathrm{Z}} \frac{\alpha_i \zeta_i}{a_i E_i^{max} \beta_i + \zeta_i}\right) - \gamma^{th}\right)^2}.$$

As stated above, the power allocation in this section follows the same procedure as in Section II-B1. Hence, the power is allocated in an iterative manner. However, the channel conditions can be such that the solution obtained is where some powers are set at their maximum constraints and some at 0. A situation can arise in this instance where $\gamma > \gamma^{th}$ due to the powers greater than their constraint being treated first in the algorithm. Hence, there should be a check at the end of the algorithm and if $\gamma > \gamma^{th}$, then the whole water-filling process is repeated, however, this time the powers which came out to be zero in the first run of the complete procedure are not included in the algorithm from the start[7].

*2) Knowledge of only Channel Statistics:* Now, we consider the case of knowledge of only statistics at the relays. As was the case in Section II-B2, the end-to-end SNR now needs to be averaged over all the links. Therefore, in this instance, the energy-efficiency optimization problem is given by

$$\min_{E_s, E_i} E_{tot}, \text{ subject to}$$
$$\bar{\gamma} \geq \gamma^{th}, \ 0 \leq E_s \leq E_s^{max}, \ 0 \leq E_i \leq E_i^{max},$$
(22)

where $\bar{\gamma}$ is given by (15). Using Appendix B and a similar argument as that in Section II-C1, it can be shown that the the optimization problem, (22), is convex. Hence, the Lagrange multiplier and other convex optimization algorithms can be applied to solve the problem. However, due to the complexity of the problem, it is very difficult to obtain closed-form expressions for the optimal solution. Therefore, we use the interior-point algorithm as utilized in Section II-B2 to obtain the optimal solution.

*3) Partial CSI:* In this section, we study the the energy-efficiency problem for the two cases of partial CSI. As was the case in Section II-B3, the both the partial CSI scenarios are special cases of the full CSI case. The objective functions and the power allocation algorithms can be obtained from the Section II-C1 for partial CSI-$\beta$ and Section II-C2 for partial CSI-$\alpha$ as was the case in Section II-B3

## III. SELECTION SCHEME

In Section II, AP relaying was considered. However, AP relaying requires additional complexity at the destination to combine the relays. Also, as the relays transmit on orthogonal channels, it consumes a huge amount of system resources and decreases throughput. To ameliorate these drawbacks of AP relaying, selective relaying has been proposed in which only the "best" relay is selected to forward the signal from the source to the destination. The selection criteria depends on the objective. For OPA, the relay which maximizes the end-to-end SNR after power allocation is selected. For energy-efficiency, the relay which minimizes the consumed energy while fulfilling the constraint on the end-to-end SNR is selected. If no relay fulfills the constraint on the end-to-end SNR, then the relay which achieves the maximum end-to-end SNR is selected.

---
[7]This was our mistake in [35]. We did not account for these special cases

## A. Optimal Power Allocation

*1) Full CSI:* For selective relaying, noting that the power is now only divided between the source and one relay, one can re-formulate the end-to-end SNR in (5), in the case that the $i$th relay is selected to transmit, as

$$\gamma_i = \eta_i E_{tot} \left( \alpha_0 + \alpha_i - \frac{\alpha_i \zeta_i}{a_i(1-\eta_i)E_{tot}\beta_i + \zeta_i} \right) \qquad i = 1, 2, \ldots, m, \tag{23}$$

where we have replaced $E_s = \eta_i E_{tot}$, $E_i = (1-\eta_i)E_{tot}$ and $0 < \eta_i \leq 1$. Ignoring the individual power constraints, the optimization problem is

$$\max_{\eta_i} \quad \eta_i E_{tot} \left( \alpha_0 + \alpha_i - \frac{\alpha_i \zeta_i}{a_i(1-\eta_i)E_{tot}\beta_i + \zeta_i} \right) \qquad i = 1, 2, \ldots, m. \tag{24}$$

The concavity of the objective function follows from the concavity of the problem for AP relaying. Taking the derivative of the objective function in (24) and equating it to zero yields the optimal solution

$$\eta_i = 1 - \frac{1}{a_i E_{tot} \beta_i} \left( \sqrt{\frac{(a_i E_{tot} \beta_i + \zeta_i)\alpha_i \zeta_i}{\alpha_0 + \alpha_i}} - \zeta_i \right), \tag{25}$$

where $\alpha_i$ and $\beta_i$ are the links associated with the $i$th relay. If $\eta_i$ is found to be such that one of the powers exceeds its individual constraints, then $\eta_i$ is adjusted that the power lies on its peak individual constraint. The case where both powers exceed their constraints is when $E_{tot} > E_s^{max} + E_i^{max}$. In this case, both the source and the selected relay transmit at their individual constraints. The algorithm for power allocation for selective relaying is

- Calculate $\eta_i$ for all the relays using (25).
- Compute the resulting $\gamma_i$ for each relay.
- Select the relay which has the maximum $\gamma_i$.

*2) Knowledge of only Channel Statistics:* As was the case for AP relaying, if CSI is not available at the relays, then the end-to-end SNR has to be averaged over the channels before power allocation is performed. Hence, now the relay which gives best performance on average is now selected. The optimization problem now, with the help of (15), is

$$\max_{\eta_i} \quad (1-\eta_i)E_{tot}\left(k_{\alpha_0}\bar{\gamma}_{\alpha_0} + k_{\alpha_0}\bar{\gamma}_{\alpha_i} - \frac{k_{\alpha_i}\bar{\gamma}_{\alpha_i}\zeta_i^{k_{\beta_i}}}{a_i^{k_{\beta_i}}\bar{\gamma}_{\beta_i}^{k_{\beta_i}}}\frac{1}{\eta_i^{k_{\beta_i}}E_{tot}^{k_{\beta_i}}}e^{\frac{\zeta_i}{a_i\eta_i E_{tot}\bar{\gamma}_{\beta_i}}}\Gamma\left(1-k_{\beta_i},\frac{\zeta_i}{a_i\eta_i E_{tot}\bar{\gamma}_{\beta_i}}\right)\right) \quad (26)$$

where now $E_s = (1-\eta_i)E_{tot}$, $E_i = \eta_i E_{tot}$ and $0 \leq \eta_i < 1$ due to ease of analysis. The concavity of the problem follows directly from the concavity of the AP case. Taking the derivative of the objective function in (26) and equating it to zero gives

$$0 = -E_{tot}\left(k_{\alpha_0}\bar{\gamma}_{\alpha_0} + k_{\alpha_0}\bar{\gamma}_{\alpha_i} - \frac{k_{\alpha_i}\bar{\gamma}_{\alpha_i}\zeta_i^{k_{\beta_i}}}{a_i^{k_{\beta_i}}E_{tot}^{k_{\beta_i}}\bar{\gamma}_{\beta_i}^{k_{\beta_i}}}\frac{1}{\eta_i^{k_{\beta_i}}}e^{\frac{\zeta_i}{a_i\eta_i E_{tot}\bar{\gamma}_{\beta_i}}}\Gamma\left(1-k_{\beta_i},\frac{\zeta_i}{a_i\eta_i E_{tot}\bar{\gamma}_{\beta_i}}\right)\right) - $$
$$(1-\eta_i)\frac{k_{\alpha_i}\bar{\gamma}_{\alpha_i}\zeta_i^{k_{\beta_i}}}{a_i^{k_{\beta_i}}E_{tot}^{k_{\beta_i}-1}\bar{\gamma}_{\beta_i}^{k_{\beta_i}}}\left(-\frac{\zeta_i}{a_i E_{tot}\bar{\gamma}_{\beta_i}}\frac{1}{\eta_i^{k_{\beta_j}+2}}\Gamma\left(1-k_{\beta_i},\frac{\zeta_i}{a_i\eta_i E_{tot}\bar{\gamma}_{\beta_i}}\right)e^{\frac{\zeta_i}{a_i\eta_i E_{tot}\bar{\gamma}_{\beta_i}}} - \right. \quad (27)$$
$$\left. k_{\beta_i}\frac{1}{\eta_i^{k_{\beta_i}+1}}\Gamma\left(1-k_{\beta_i},\frac{\zeta_i}{a_i\eta_i E_{tot}\bar{\gamma}_{\beta_i}}\right)e^{\frac{\zeta_i}{a_i\eta_i E_{tot}\bar{\gamma}_{\beta_i}}} + \frac{\zeta_i^{-k_{\beta_i}+1}}{a_i^{-k_{\beta_i}+1}\eta_i^2 E_{tot}^{-k_{\beta_i}+1}\bar{\gamma}_{\beta_i}^{-k_{\beta_i}+1}}\right).$$

Equation (27) can be solved numerically using algorithms such as bisection, Newton's method etc to yield the optimal value of $\eta_i$. Similar to the full CSI case, $\eta_i$ is found for all the relays and then the relay which maximizes the averaged end-to-end SNR is selected. For the special case of Rayleigh fading, $\eta_i$ can be obtained from

$$0 = -E_{tot}\left(\bar{\gamma}_{\alpha_0} + \bar{\gamma}_{\alpha_i} - \frac{\zeta_i\bar{\gamma}_{\alpha_i}}{a_i\eta_i E_{tot}\bar{\gamma}_{\beta_i}}e^{\frac{\zeta_i}{a_i\eta_i E_{tot}\bar{\gamma}_{\beta_i}}}E_1\left(\frac{\zeta_i}{a_i\eta_i E_{tot}\bar{\gamma}_{\beta_i}}\right)\right) - (1-\eta)\frac{\zeta_i\bar{\gamma}_{\alpha_i}}{a_i\bar{\gamma}_{\beta_i}}$$
$$\left(-\frac{1}{\eta_i^2}e^{\frac{\zeta_i}{a_i\eta_i E_{tot}\bar{\gamma}_{\beta_i}}}E_1\left(\frac{\zeta_i}{a_i\eta_i E_{tot}\bar{\gamma}_{\beta_i}}\right) + \frac{1}{\eta_i}\left(\frac{1}{\eta_i} - E_1\left(\frac{\zeta_i}{a_i\eta_i E_{tot}\bar{\gamma}_{\beta_i}}\right)e^{\frac{\zeta_i}{a_i\eta_i E_{tot}\bar{\gamma}_{\beta_i}}}\frac{\zeta_i}{a_i\eta_i^2 E_{tot}\bar{\gamma}_{\beta_i}}\right)\right). \quad (28)$$

*3) Partial CSI:* The two partial CSI cases can again be handled in the same way as in Section II-B3. The objective functions and the power allocation algorithms can be obtained from the full CSI case in Section III-A1 and the channel statistics case in Section III-A2 in the manner as outlined in Section II-B3.

*B. Energy-Efficiency*

*1) Full CSI:* In the case of full CSI, the energy-efficiency problem for the $i$th selected relay is

$$\min_{Es,Ei} E_{tot}, \text{ subject to}$$
$$\gamma_i \geq \gamma^{th}, \ 0 \leq E_s \leq E_s^{max}, \ 0 \leq E_i \leq E_i^{max}, \quad (29)$$

where

$$\gamma_i = E_s \left( \alpha_0 + \alpha_i - \frac{\alpha_i \zeta_i}{a_i E_i \beta_i + \zeta_i} \right). \tag{30}$$

The optimization problem, (29), is solved for all the relays and the relay which minimizes $E_{tot}$ while fulfilling the constraint on $\gamma_i$ is selected. If no relay fulfills the constraint on $\gamma_i$, then the relay which maximizes $\gamma_i$ is selected.

Ignoring the individual constraints and taking advantage of the fact that at the optimal solution $\gamma = \gamma^{th}$, we can write

$$E_s = \frac{a_i E_i \beta_i \gamma^{th} + \zeta_i \gamma^{th}}{\alpha_0 \zeta_i + a_i E_i \alpha_0 \beta_i + a_i E_i \alpha_i \beta_i}. \tag{31}$$

Thus, ignoring the individual constraints, (29) can be re-written as

$$\min_{E_i} \quad \frac{a_i E_i \beta_i \gamma^{th} + \zeta_i \gamma^{th}}{\alpha_0 \zeta_i + a_i E_i \alpha_0 \beta_i + a_i E_i \alpha_i \beta_i} + E_i. \tag{32}$$

Taking the derivative and equating it to 0 gives $E_i$ as

$$E_i = \frac{\sqrt{a_i \alpha_i \beta_i \zeta_i \gamma^{th}} - \alpha_0 \zeta_i}{a_i \beta_i (\alpha_0 + \alpha_i)}. \tag{33}$$

Substituting (33) in (31) yields

$$E_s = \frac{\left( \sqrt{a_i \alpha_i \beta_i \zeta_i} + \zeta_i \alpha_0 \right) \gamma^{th}}{(\alpha_0 + \alpha_i) \sqrt{a_i \alpha_i \beta_i \zeta_i \gamma^{th}}}. \tag{34}$$

Incorporating the individual constraints gives the water-filling solution

$$E_s = \left( \frac{\left( \sqrt{a_i \alpha_i \beta_i \zeta_i} + \zeta_i \alpha_0 \right) \gamma^{th}}{(\alpha_0 + \alpha_i) \sqrt{a_i \alpha_i \beta_i \zeta_i \gamma^{th}}} \right)_0^{E_s^{max}} \tag{35}$$

$$E_i = \left( \frac{\sqrt{a_i \alpha_i \beta_i \zeta_i \gamma^{th}} - \alpha_0 \zeta_i}{a_i \beta_i (\alpha_0 + \alpha_i)} \right)_0^{E_i^{max}}. \tag{36}$$

*2) Knowledge of only Channel Statistics:* In this case, the selection procedure and the optimization problem are the same as in Section III-B1, however the constraint on the end-to-end SNR changes to $\bar{\gamma}_i \geq \gamma^{th}$, where

$$\bar{\gamma}_i = E_s \left( k_{\alpha_0} \bar{\gamma}_{\alpha_0} + k_{\alpha_i} \bar{\gamma}_{\alpha_i} - \frac{k_{\alpha_i} \bar{\gamma}_{\alpha_i} \zeta_i^{k_{\beta_i}}}{a_i^{k_{\beta_i}} \bar{\gamma}_{\beta_i}^{k_{\beta_i}}} \frac{1}{E_i^{k_{\beta_i}}} e^{\frac{\zeta_i}{a_i E_i \bar{\gamma}_{\beta_i}}} \Gamma \left( 1 - k_{\beta_i}, \frac{\zeta_i}{a_i E_i \bar{\gamma}_{\beta_i}} \right) \right). \tag{37}$$

Again exploiting the equality of the on $\bar{\gamma}_i$, we obtain

$$E_s = \frac{\gamma^{th} E_i^{k_{\beta_i}}}{\left(k_{\alpha_0}\bar{\gamma}_{\alpha_0} + k_{\alpha_i}\bar{\gamma}_{\alpha_i}\right) E_i^{k_{\beta_i}} - \frac{k_{\alpha_i}\bar{\gamma}_{\alpha_i}\zeta_i^{k_{\beta_i}}}{a_i^{k_{\beta_i}}\bar{\gamma}_{\beta_i}^{k_{\beta_i}}} e^{\frac{\zeta_i}{a_i E_i \bar{\gamma}_{\beta_i}}} \Gamma\left(1 - k_{\beta_i}, \frac{\zeta_i}{a_i E_i \bar{\gamma}_{\beta_i}}\right)}. \tag{38}$$

Therefore, the optimization problem becomes

$$\min_{E_i} \frac{\gamma^{th} E_i^{k_{\beta_i}}}{\left(k_{\alpha_0}\bar{\gamma}_{\alpha_0} + k_{\alpha_i}\bar{\gamma}_{\alpha_i}\right) E_i^{k_{\beta_i}} - \frac{k_{\alpha_i}\bar{\gamma}_{\alpha_i}\zeta_i^{k_{\beta_i}}}{a_i^{k_{\beta_i}}\bar{\gamma}_{\beta_i}^{k_{\beta_i}}} e^{\frac{\zeta_i}{a_i E_i \bar{\gamma}_{\beta_i}}} \Gamma\left(1 - k_{\beta_i}, \frac{\zeta_i}{a_i E_i \bar{\gamma}_{\beta_i}}\right)} + E_i. \tag{39}$$

Taking the derivative and equating to 0 gives

$$\left(\left(k_{\alpha_0}\bar{\gamma}_{\alpha_0} + k_{\alpha_i}\bar{\gamma}_{\alpha_i}\right) E_i^{k_{\beta_i}} - \frac{k_{\alpha_i}\bar{\gamma}_{\alpha_i}\zeta_i^{k_{\beta_i}}}{a_i^{k_{\beta_i}}\bar{\gamma}_{\beta_i}^{k_{\beta_i}}} e^{\frac{\zeta_i}{a_i E_i \bar{\gamma}_{\beta_i}}} \Gamma\left(1 - k_{\beta_i}, \frac{\zeta_i}{a_i E_i \bar{\gamma}_{\beta_i}}\right)\right)^2 - \frac{\gamma^{th} k_{\alpha_i} k_{\beta_i} \bar{\gamma}_{\alpha_i}\zeta_i^{k_{\beta_i}}}{a_i^{k_{\beta_i}}\bar{\gamma}_{\beta_i}^{k_{\beta_i}}} e^{\frac{\zeta_i}{a_i E_i \bar{\gamma}_{\beta_i}}} \times$$

$$\Gamma\left(1 - k_{\beta_i}, \frac{\zeta_i}{a_i E_i \bar{\gamma}_{\beta_i}}\right) E_i^{k_{\beta_i}-1} + \frac{\gamma^{th} k_{\alpha_i} \bar{\gamma}_{\alpha_i}\zeta_i}{a_i E_i^2 \bar{\gamma}_{\beta_i}} - \frac{\gamma^{th} k_{\alpha_i} \bar{\gamma}_{\alpha_i}\zeta_i^{k_{\beta_i}+1}}{a_i^{k_{\beta_i}+1}\bar{\gamma}_{\beta_i}^{k_{\beta_i}+1}} \Gamma\left(1 - k_{\beta_i}, \frac{\zeta_i}{a_i E_i \bar{\gamma}_{\beta_i}}\right) e^{\frac{\zeta_i}{a_i E_i \bar{\gamma}_{\beta_i}}} E_i^{k_{\beta_i}-2} = 0. \tag{40}$$

The above equation can be solved through bisection search to yield the value of $E_i$ which can be substituted back into (38) to obtain $E_s$. The maximum of $E_s$ and $E_i$ is checked and if it exceeds its peak constraint, then it is set at it its peak constraint and the other power is obtained from the constraint. If no power exceeds its respective peak constraint, then the minimum power is checked and if it is below 0, it is set to zero and the other power is obtained from the constraint.

For Rayleigh fading, (41) simplifies to

$$\left(\left(\bar{\gamma}_{\alpha_0} + \bar{\gamma}_{\alpha_i}\right) E_i - \frac{\bar{\gamma}_{\alpha_i}\zeta_i}{a_i\bar{\gamma}_{\beta_i}} e^{\frac{\zeta_i}{a_i E_i \bar{\gamma}_{\beta_i}}} E_1\left(\frac{\zeta_i}{a_i E_i \bar{\gamma}_{\beta_i}}\right)\right)^2 - \frac{\gamma^{th}\bar{\gamma}_{\alpha_i}\zeta_i}{a_i\bar{\gamma}_{\beta_i}} e^{\frac{\zeta_i}{a_i E_i \bar{\gamma}_{\beta_i}}} E_1\left(\frac{\zeta_i}{a_i E_i \bar{\gamma}_{\beta_i}}\right)$$
$$+ \frac{\gamma^{th}\bar{\gamma}_{\alpha_i}\zeta_i}{a_i E_i^2 \bar{\gamma}_{\beta_i}} - \frac{\gamma^{th}\bar{\gamma}_{\alpha_i}\zeta_i}{a_i E_i \bar{\gamma}_{\beta_i}} E_1\left(\frac{\zeta_i}{a_i E_i \bar{\gamma}_{\beta_i}}\right) e^{\frac{\zeta_i}{a_i E_i \bar{\gamma}_{\beta_i}}} = 0. \tag{41}$$

*3) Partial CSI:* The two partial CSI cases can be obtained from the full CSI case and the channel statistics case as outlined in Section II-C3.

## IV. NUMERICAL RESULTS

We present numerical results to the schemes discussed in this section. For the numerical results, all the noise variances are taken to be equal, i.e. $\sigma_{sd}^2 = \sigma_{si}^2 = \sigma_{id}^2 = \sigma^2$. The average SNR of all the links are set at 0.5, i.e. $\bar{\gamma}_{\alpha_i} = \bar{\gamma}_{\beta_i} = 0.5$, for $i$. All the shape parameters are taken to be 1 except when indicated. The results shown are for a system with 3 relays ($m = 3$). The relay gain for all relay is taken to be 1. The peak individual constraints are set as $E_s^{max} = 3$ and $E_i^{max} = 3$ for all $i$. For OPA, $E_{tot}$ is taken to be 5.5 and for energy-efficiency, $\gamma^{th}$ is taken to be 5 dB. Also, for the energy-efficiency problem, it might be the case that due to channel conditions the constraint on the end-to-end SNR cannot be met. In that case all transmitting relays and the source transmit at their maximum power.

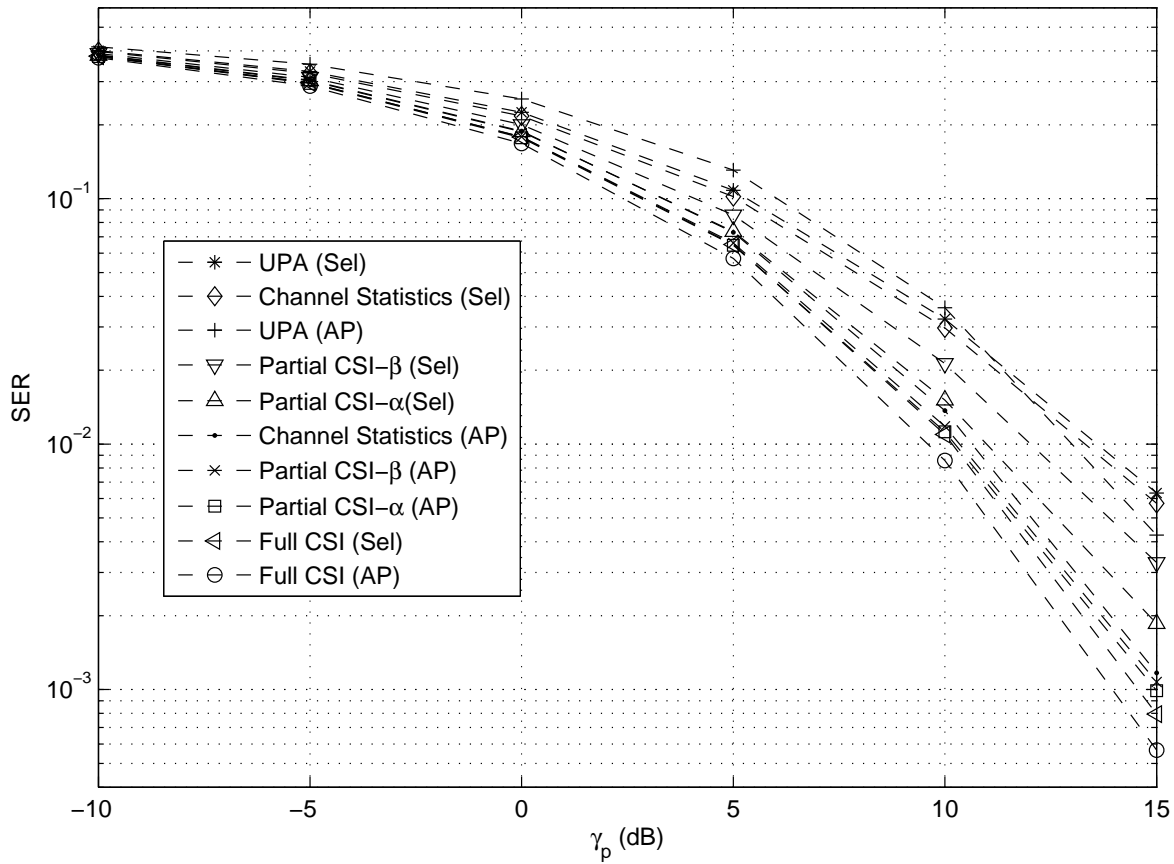

Fig. 1: Comparison of SER performance.

Figure 1 shows the comparison of the SER for the different schemes for binary PSK as a function of $\gamma_p$[8]. Firstly, we compare among the AP and selection schemes for the different cases of CSI at the relays. It is evident from Figure 1 that the four OPA AP schemes comfortably provide better performance than uniform power allocation (UPA(AP)). Full CSI (AP) gives the greatest gain, as one would expect, of around 3.4 dB over UPA (AP) at a SER of $10^{-2}$, while partial CSI-$\alpha$ (AP), partial CSI-$\beta$ (AP) and channel statistics (AP) display a gain of around 2.8 dB, 2.6 dB and 2.4 dB, respectively at the same SER. As can be noted from the gains of the four OPA schemes over UPA (AP), partial CSI-$\alpha$ (AP) outperforms partial CSI-$\beta$ (AP) by a slim margin. However, it must be remembered partial CSI-$\alpha$ (AP) requires knowledge of an additional link, namely the source-destination link. Moreover, at a SER of $10^{-2}$, all the four OPA AP schemes lie within a range of approximately 1.1 dB. Coming to the four OPA schemes for selective relaying. As expected, the pattern is similar to the AP case, the four OPA schemes perform better than UPA (Sel)[9]. However, the difference in performance for the four OPA selective relaying schemes is large as compared to the difference in OPA AP case. Full CSI (Sel) provides a gain of almost 0.8 dB over partial CSI-$\alpha$ (Sel) at a SER of $10^{-2}$ while partial CSI-$\alpha$ (Sel) in turn gives a gain of around 1 dB over partial CSI-$\beta$ (Sel). Similarly, Channel Statistics (Sel) trails partial CSI-$\beta$ (Sel) by 1.3 dB. However, Channel Statistics (Sel) only provides a gain of 0.2 dB over UPA (Sel). Thus, unlike AP, in selective relaying, there is quite a large difference in performance of the four OPA schemes among themselves and a small difference in performance among them and their UPA counterparts. Now comparing AP and selective relaying, the difference between AP and selective relaying increases with less CSI the system has. In the OPA Full CSI case, the difference between AP and selective relaying is only 0.6 dB, however it increases with the decrease in channel knowledge and for knowledge of channel statistics only it reaches 2.6 dB. Another interesting point to note is that UPA (Sel) outperforms UPA (AP) at low values of $\gamma_p$[10].

---

[8]$\gamma_p = \frac{E_{tot}}{\sigma^2}$

This is due to the fact that even though AP has more relays, the total power is the same for both AP and selective relaying. For UPA (AP), this power is equally distributed among the 3 relay and the source, however, for selective relaying the power is shared between only two nodes and moreover, the relay which maximizes the end-to-end SNR is selected. Thus, more power allocated to the relay which has better channel conditions and UPA (Sel) performs better than UPA (AP). However, as $\gamma_p$ increases, all the relays see good channel conditions, in general, and the gain of AP is seen.

The throughputs of all the OPA schemes are shown in Figure 2. All the selective relaying schemes give better throughput than all the AP schemes. This is due to the orthogonal distribution of sources in AP.

---

[9]For selective relaying, each case of CSI will have a different UPA scheme because of the different selection criterias. However, as to not make the figure cluttered we only show the case for channel statistics. All the other UPA selection schemes show a similar behaviour to their corresponding optimal schemes as UPA for channel statistics exhibits with Channel Statistics (Sel). Hence, UPA (Sel) refers to UPA for channel statistics.

[10]It must be kept in mind that UPA (Sel) refers to the case of knowledge of channel statistics only. Hence UPA for selective relaying, which has close to Full CSI (Sel performance as mentioned previously), outperforms UPA (AP) at moderate and high values of $\gamma_p$.

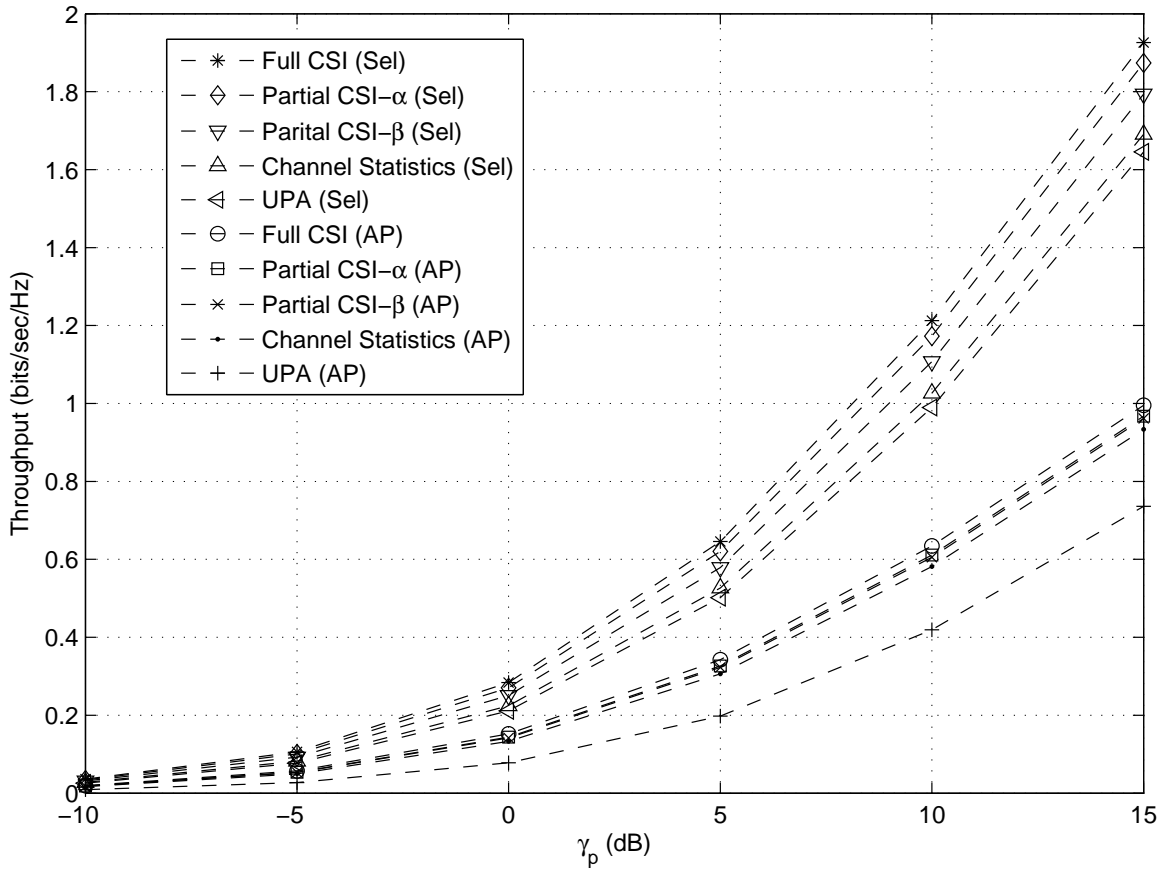

Fig. 2: Comparison of Throughput.

Hence, transmitting one packet of information requires $m+1$ time slots for AP while it requires only 2 slots for selection. Furthermore, the same pattern is observed for the throughputs as was seen for the SER for the CSI case. Partial CSI-$\alpha$ outperforms Partial CSI-$\beta$ and knowledge of only channel statistics gives the worst performance.

Figures 3 and 4 show the performance of AP and selective relaying for the energy-efficiency problem, respectively. We will first discuss energy-efficiency for AP relaying by considering Figure 3 and then move onto Figure 4 for selective relaying. In the end, we will compare AP and selective relaying. Now, Figure 3(a) shows the total power consumed, $E_{tot}$, as a function of $\gamma_s$[11] and Figure 3(b) show $R$[12] as a function of $\gamma_s$. Figure 3(a) shows that Full CSI (AP) consumes the least amount of energy at low $\gamma_s$, however, as $\gamma_s$ increases, Full CSI (AP) consumes the most amount of energy. This is a little counter intuitive as one

---

[11] $\gamma_s = \frac{1}{\sigma^2}$.
[12] $R = \log_2(1+\gamma)$. Hence, $R$ is a linearly increasing function of $\gamma$.

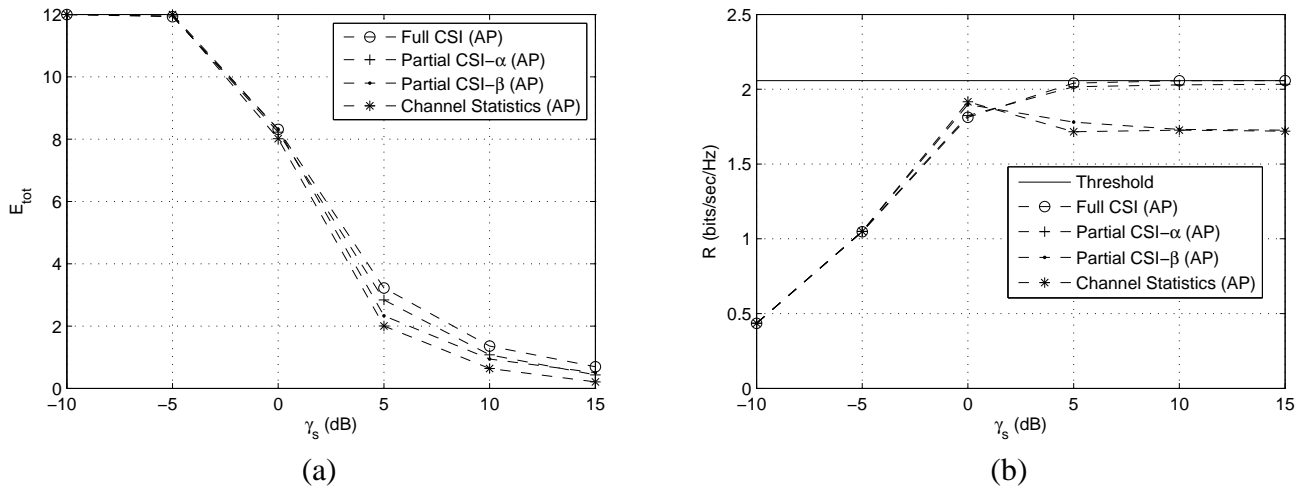

Fig. 3: Energy-efficiency for AP (a) $E_{tot}$ (b) $R$.

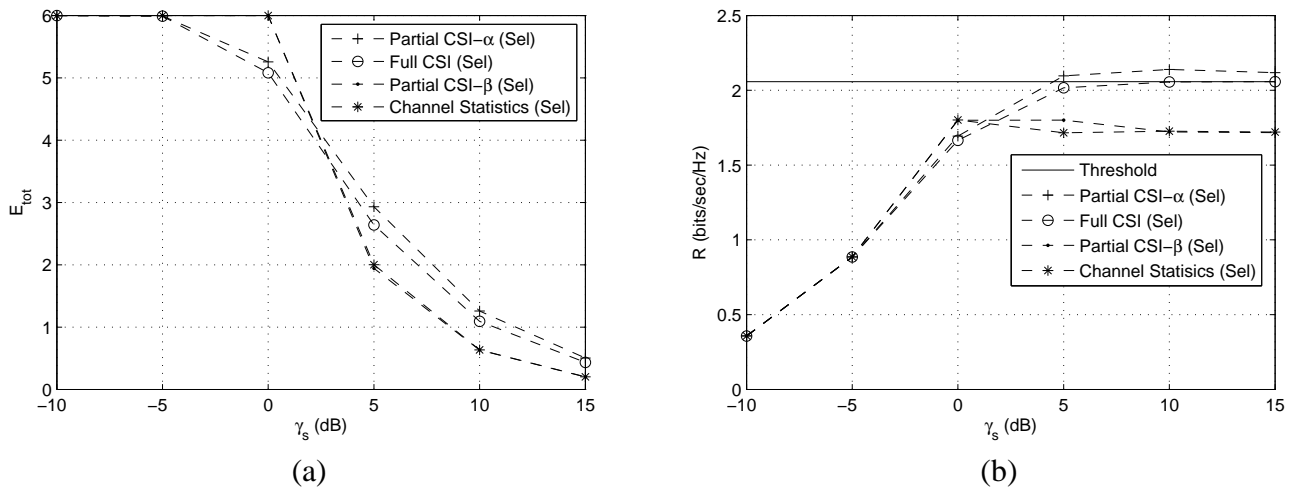

Fig. 4: Energy-efficiency for selective relaying (a) $E_{tot}$ (b) $R$.

would expect Full CSI (AP) to give the best performance. However, the analysis is incomplete without looking at the achieved end-to-end SNR which is shown by 3(b). The threshold line represents represents the value of $R$ for the constraint on the end-to-end SNR, i.e. threshold$= \log_2(1 + \gamma^{th})$. As one can see from Figure 3(b) that even though Full CSI (AP) consumes the most energy at high $\gamma_s$, it also achieves the threshold constraint on average. Moreover, at low $\gamma_s$, Full CSI (AP) consumes a little less energy than all the other CSI cases and achieves the same $R$ on average. The most energy consumed after Full CSI (AP) is by Partial CSI-$\alpha$ (AP), however it also provides the second best $R$ after the full CSI case. Partial CSI-$\beta$ (AP) and Channel Statistics (AP) achieve the worst $R$. Selective relaying exhibits a similar behaviour in Figure 4 with one major difference. At high $\gamma_s$, Partial CSI-$\alpha$ (Sel) achieves a greater $R$ than Full CSI (Sel) which approximately achieves the threshold value. However, this comes at increased energy consumption as shown by Figure 4(b) due to the mismatch in instantaneous values and the average

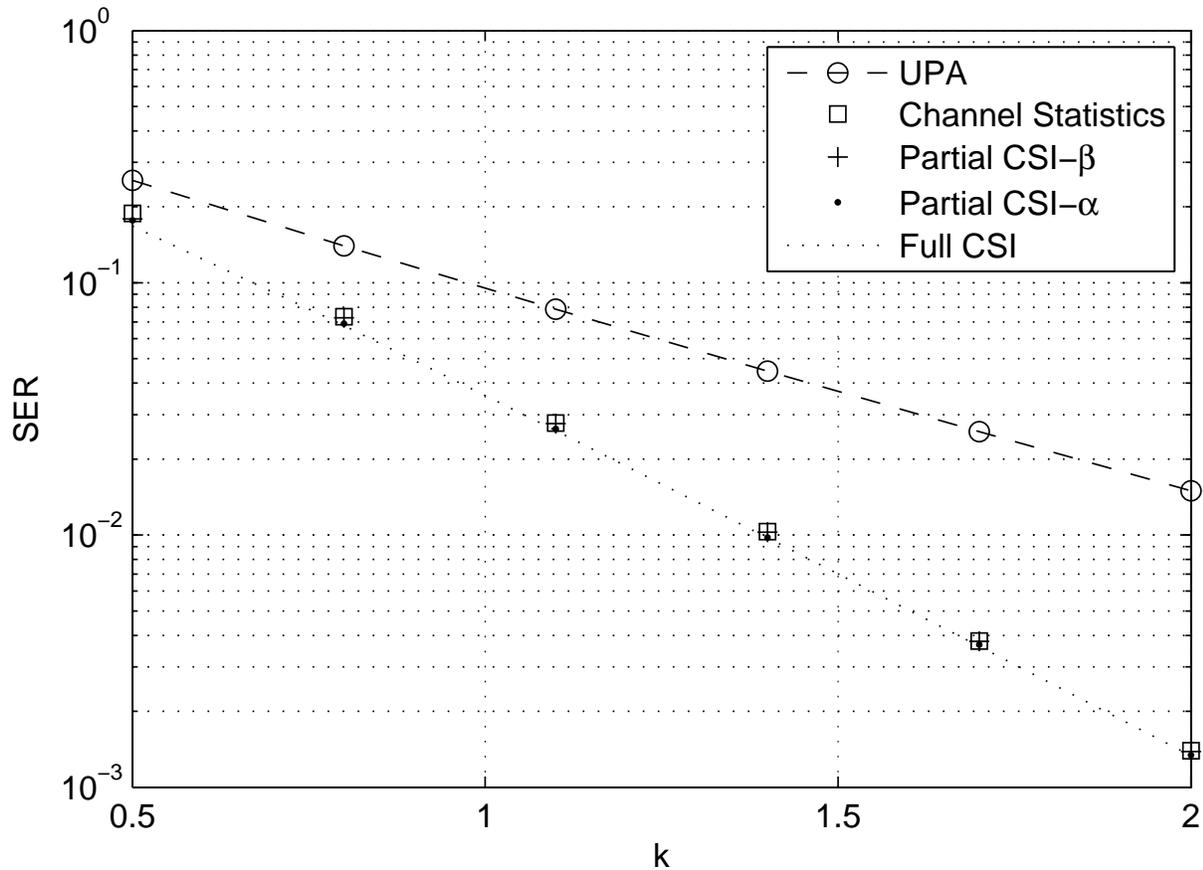

Fig. 5: Effect of shaping parameter.

value.

Now comparing AP relaying and selective relaying, all the selective relaying cases consume less energy than their AP counterparts for low values of $\gamma_s$ which is understandable as when the channel conditions are bad, AP will consume 12 energy units while selection only 6 due to only one relay being active. Consequently, the selective relaying CSI cases achieve a lower $R$ than their corresponding AP cases. However, this pattern continues even for at high $\gamma_s$ where AP would expect to consume less energy. The reason for this is again the case when the channel conditions are bad and the constraint on the end-to-end SNR cannot be met. Even though bad channel conditions occur rarely for at high $\gamma_s$, but, in these instances, AP consumes twice the power of selective relaying which pushes the average power consumed for AP a little above than that of selective relaying. Therefore, it seems that for energy-efficiency, selective relaying is a better option than AP even at high $\gamma_s$.

The effect of the shaping parameter $k$[13] is shown in Figure 5[14] for $\gamma_p = 0$ dB. As can be seen from Figure 5, a higher shape parameter means better performance for all the CSI cases. However, this does not effect the difference in the three CSI cases.

## V. CONCLUSIONS

We have studied power allocation to maximize the end-to-end SNR under a total power constraint and to minimize the the total power consumed while maintaining the end-to-end SNR above a required threshold for a fixed-gain AF relay network. We have studied both problems for the relay network operating in AP mode where all the relays participate in signal forwarding and operating in selection mode in which only the selected relay forwards the signal to the destination. Furthermore, we have also considered the cases of full CSI, partial CSI and knowledge of only channel statistics at the relay for both modes of operation and for both optimization problems. We demonstrate the gain achieved by allocating power optimally over UPA. We also give insight into the performance of the system for both problems, for both AP relaying and selective relaying and for all three cases of CSI at the relays. Additionally, we also develop inequalities in Appendix B which may prove to be useful in future works. For these reasons, we believe that our work is a valuable contribution to the already available literature on power allocation strategies for fixed-gain AF relays.

## APPENDIX A

Writing down the objective function

$$\gamma = E_s \left( \sum_{i=0}^{m} \alpha_i - \sum_{i=1}^{m} \frac{\alpha_i \zeta_i}{a_i E_i \beta_i + \zeta_i} \right). \tag{42}$$

The objective function, in general, not convex and concave. However, as we show below, it is concave (its negative is convex) for the domain we are interested in.

Define vector $\mathtt{E}$ as

$$\mathtt{E} = [E_s\ E_1\ E_2\ \cdots\ E_m]^T \qquad \mathtt{E} \succ \mathbf{0}. \tag{43}$$

---

[13]$k_{\alpha_i} = k_{\beta_i} = k$ for all $i$

[14]We only show the results for AP as selection also displays similar results and behaviour and they are omitted so as to not clutter the plot and make it difficult to follow.

Now let us define

$$f(\mathbf{E}) = [1\ 0\ 0\ \ldots 0]\mathbf{E} = E_s \quad g(\mathbf{E}) = \sum_{i=0}^{m} \alpha_i - \sum_{i=1}^{m} \frac{\alpha_i \zeta_i}{a_i E_i \beta_i + \zeta_i}. \tag{44}$$

Both $f$ and $g$ are positive and increasing on their domain. For $f$ to be concave

$$f(\theta x + (1-\theta)y) \geq \theta f(x) + (1-\theta)f(y), \tag{45}$$

where $0 \leq \theta \leq 1$. The left hand side (LHS) in the above is $\theta x_1 + (1-\theta)y_1$ and the right hand side (RHS) is equal to $\theta x_1 + (1-\theta)y_1$. As the LHS is equal to the RHS, $f$ is concave. To show that $g$ is concave, forming the Hessian

$$\mathbf{H}_g = \begin{bmatrix} 0 & 0 & 0 & \cdots & \cdots & 0 \\ 0 & -\frac{E_s \alpha_1 \zeta_1 a_1^2 \beta_1^2}{(a_1 E_1 \beta_1 + \zeta_1)^3} & 0 & \cdots & \cdots & 0 \\ 0 & 0 & -\frac{E_s \alpha_2 \zeta_2 a_2^2 \beta_2^2}{(a_2 E_2 \beta_2 + \zeta_2)^3} & 0 & \cdots & 0 \\ \vdots & \vdots & \ddots & \ddots & \ddots & \vdots \\ \vdots & \vdots & \ddots & \ddots & \ddots & \vdots \\ 0 & 0 & \cdots & \cdots & \cdots & -\frac{E_s \alpha_m \zeta_m a_m^2 \beta_m^2}{(a_m E_m \beta_m + \zeta_m)^3} \end{bmatrix}. \tag{46}$$

As the eigenvalues of $\mathbf{H}_g$ are non-negative, $\mathbf{H}_g$ is negative semi-definite, and hence $g$ is concave. Now let us define

$$h(\mathbf{E}) = f(\mathbf{E})g(\mathbf{E}) = E_s \left( \sum_{i=0}^{m} \alpha_i - \sum_{i=1}^{m} \frac{\alpha_i \zeta_i}{a_i E_i \beta_i + \zeta_i} \right). \tag{47}$$

For $h$ to be concave ($-h$ to be convex)

$$h(\theta x + (1-\theta)y) \geq \theta h(x) + (1-\theta)h(y). \tag{48}$$

Therefore for concavity we have to show

$$\Delta \leq 0, \tag{49}$$

where

$$\Delta = \theta(fg)(x) + (1-\theta)(fg)(y) - (fg)(\theta x + (1-\theta)y) \tag{50}$$

As $f$ and $g$ are both positive and concave

$$(fg)(\theta x + (1-\theta)y) \geq (\theta f(x) + (1-\theta)f(y))(\theta g(x) + (1-\theta)g(y)) \tag{51}$$

Substituting (51) in the expression of $\Delta$ one has

$$\Delta \leq \theta(fg)(x) + (1-\theta)(fg)(y) - (\theta f(x) + (1-\theta)f(y))(\theta g(x) + (1-\theta)g(y)). \tag{52}$$

$$\Delta \leq \theta f(x)g(x) + (1-\theta)f(y)g(y) - \theta^2 f(x)g(x) - (1-\theta)^2 f(y)g(y) \\ - \theta(1-\theta)f(x)g(y) - \theta(1-\theta)f(y)g(x). \tag{53}$$

After some manipulation

$$\Delta \leq \theta(1-\theta)D(x,y), \tag{54}$$

where $D(x,y) = (f(x) - f(y))(g(x) - g(y))$. If $D(x,y) \leq 0$, then the proof of concavity is complete. For optimal power allocation, if $f(x) > f(y)$ then $\sum_{i \in y} E_i > \sum_{i \in x} E_i$ from the total power constraint. Therefore, with power allocation, $g(y) > g(x)$, as it $g(.)$ is a concave functions of the relay powers, implying $D(x,y) < 0$ and hence, concavity. Similarly, if $g(x) > g(y)$, then power allocation means that $\sum_{i \in x} E_i > \sum_{i \in y} E_i$, which in turns means $f(y) > f(x)$. Thus, $D(x,y) < 0$ and the objective function is concave.

## APPENDIX B

Writing down the objective function

$$\bar{\gamma} = E_s \sum_{i=0}^{m} k_{\alpha_i} \bar{\gamma}_{\alpha_i} - E_s \sum_{i=1}^{m} \frac{k_{\alpha_i} \bar{\gamma}_{\alpha_i} \zeta_i^{k_{\beta_i}}}{a_i^{k_{\beta_i}} \bar{\gamma}_{\beta_i}^{k_{\beta_i}}} \frac{1}{E_i^{k_{\beta_i}}} e^{\frac{\zeta_i}{a_i E_i \bar{\gamma}_{\beta_i}}} \Gamma\left(1 - k_{\beta_i}, \frac{\zeta_i}{a_i E_i \bar{\gamma}_{\beta_i}}\right)$$

It is obvious that it is a concave function of $E_s$. To check concavity with respect to $E_j$, we find the second derivative

$$\frac{\partial^2 \bar{\gamma}}{\partial E_j^2} = -\frac{E_s k_{\alpha_j} \bar{\gamma}_{\alpha_j} \zeta_j^{k_{\beta_j}}}{a_j^{k_{\beta_j}} E_j^{k_{\beta_j}+2} \bar{\gamma}_{\beta_j}^{k_{\beta_j}}} S(v, k_{\beta_j}), \tag{55}$$

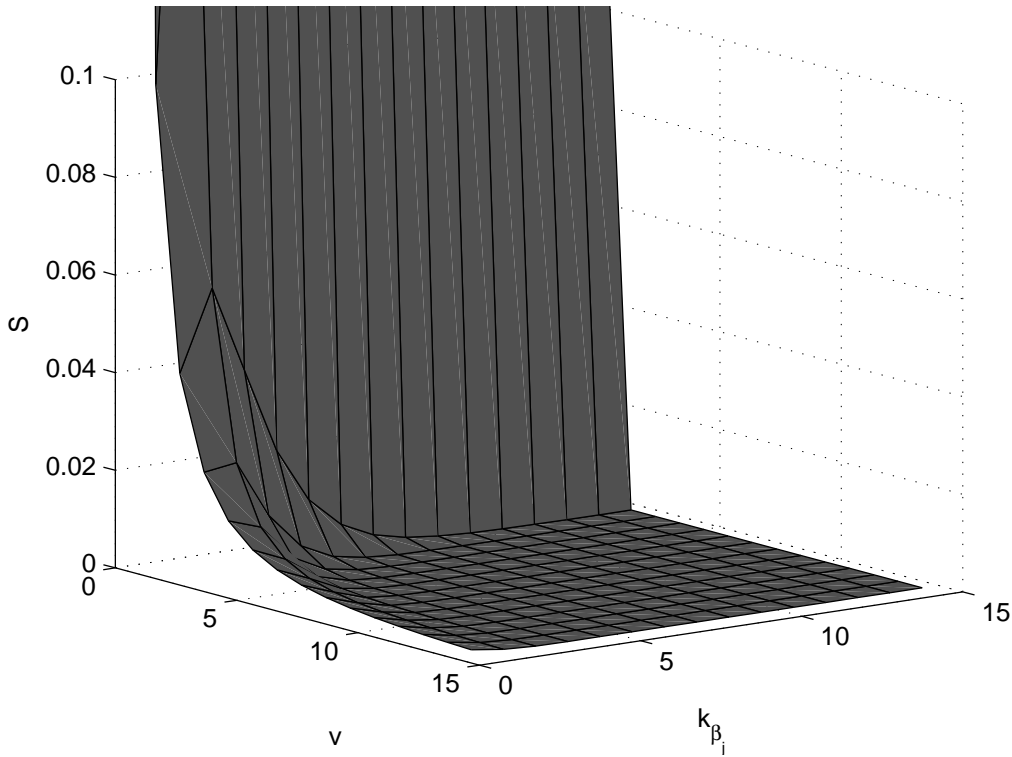

Fig. 6: Plot of $S(v, k_{\beta_j})$.

where $v = \frac{\zeta_j}{a_j E_j \bar{\gamma}_{\beta_j}}$ and

$$S(v, k_{\beta_j}) = -(2 + k_{\beta_j})v^{-k_{\beta_j}+1} - v^{-k_{\beta_j}+2} + \Gamma\left(1 - k_{\beta_j}, v\right) e^v \left(v^2 + 2k_{\beta_j}v + 2v + k_{\beta_j}^2 + k_{\beta_j}\right) \quad (56)$$

Therefore, for concavity with respect to $E_j$, we have to show

$$S(v, k_{\beta_j}) \geq 0 \qquad v > 0, \ k_{\beta_j} > 0. \quad (57)$$

Figure 6 shows the plot of $S(.)$. It is apparent from Figure 6 that $S(.) \geq 0$. However, we did not show $S(.)$ for very low and high values of $v$ and $k_{\beta_j}$). The is due to the fact that for low values fo the parameters, $S(.)$ has a very high value exceeding $10^{20}$ as can be seen from the pattern of Figure 6. At very high values of $v$ and $k_{\beta_j}$), the values of $S(.)$ become very low even falling below $10^{-18}$ after which software packages like MATLAB can't accurately characterize the number on plots. However, the value does not become negative at any point. Therefore, the objective function is concave with respect to $E_j$. Now using a similar argument as in Appendix A, it can be shown that the objective function jointly concave with respect to $E_s$ and $E_j$.

For the special case of Rayleigh fading, the second derivative becomes

$$\frac{\partial^2 \bar{\gamma}}{\partial E_j^2} = -E_s \frac{\zeta_j \bar{\gamma}_{\alpha_j}}{a_j E_j^3 \bar{\gamma}_{\beta_j}} \left( -3 - \frac{\zeta_j}{a_j E_j \bar{\gamma}_{\beta_j}} + E_1 \left( \frac{\zeta_j}{a_j E_j \bar{\gamma}_{\beta_j}} \right) e^{\frac{\zeta_j}{a_i E_j \bar{\gamma}_{\beta_j}}} \frac{\zeta_j^2}{a_j^2 E_j^2 \bar{\gamma}_{\beta_j}^2} \right.$$
$$\left. + 2 E_1 \left( \frac{\zeta_j}{a_j E_j \bar{\gamma}_{\beta_j}} \right) e^{\frac{\zeta_j}{a_i E_j \bar{\gamma}_{\beta_j}}} + 4 E_1 \left( \frac{\zeta_j}{a_j E_j \bar{\gamma}_{\beta_j}} \right) e^{\frac{\zeta_j}{a_i E_j \bar{\gamma}_{\beta_j}}} \frac{\zeta_j}{a_j E_j \bar{\gamma}_{\beta_j}} \right). \tag{58}$$

In this case, the proof of concavity establishes the relationship

$$E_1(v) e^v (v^2 + 2 + 4v) - v - 3 > 0 \qquad v > 0. \tag{59}$$